\documentclass{aa}

\input psfig.sty

\begin{document}
\newcommand{\mo}{$M_{\odot}$}
\newcommand{\mob}{$M_{\odot}\;$}

\title{Habitable Zones and UV Habitable Zones around Host Stars}

\author{Jianpo Guo\inst{1,2,3}
        \and Fenghui Zhang\inst{1,2}
        \and Xianfei Zhang\inst{1,2,3}
        \and Zhanwen Han\inst{1,2}
        }

\offprints{Guo et al.}

\institute{National Astronomical Observatories/Yunnan Observatory,
Chinese Academy of Sciences,
           Kunming, 650011, P.R. China\\
           \email{guojianpo1982@hotmail.com}
          \and
          Key Laboratory for the Structure and Evolution of
Celestial Objects, Chinese Academy of Sciences,
           Kunming, 650011, P.R. China\\
           \and
           Graduate School of the Chinese Academy of Sciences,
Beijing, 100049, P.R. China\\
          }

\date{Received 13 Aug. 2009; accepted 11 Oct. 2009}

\abstract{Ultraviolet radiation is a double-edged sword to life. If
it is too strong, the terrestrial biological systems will be
damaged. And if it is too weak, the synthesis of many biochemical
compounds can not go along. We try to obtain the continuous
ultraviolet habitable zones, and compare the ultraviolet habitable
zones with the habitable zones of host stars. Using the boundary
ultraviolet radiation of ultraviolet habitable zone, we calculate
the ultraviolet habitable zones of host stars with masses from 0.08
to 4.00 \mo. For the host stars with effective temperatures lower
than 4,600 K, the ultraviolet habitable zones are closer than the
habitable zones. For the host stars with effective temperatures
higher than 7,137 K, the ultraviolet habitable zones are farther
than the habitable zones. For hot subdwarf as a host star, the
distance of the ultraviolet habitable zone is about ten times more
than that of the habitable zone, which is not suitable for life
existence.

}

\titlerunning{HZs and UV-HZs}

\authorrunning{Guo et al.}

\maketitle

\section{Introduction}
Typically, stellar habitable zone (HZ) is defined as a region near
the host star where water at the surface of a terrestrial planet is
in liquid phase, which has been widely researched (eg., Hart, 1978;
Kasting et al., 1993; Franck et al., 2000; Noble et al., 2002). The
boundary flux of HZ not only depends on luminosity, but also depends
on effective temperature ($T_{\rm eff}$) (eg., Forget \&
Pierrehumbert, 1997; Williams \& Kasting, 1997; Mischna et al.,
2000; Jones, 2004; Jones et al., 2006). As the higher $T_{\rm eff}$,
the less the infrared fraction in luminosity, and the less this
fraction, the less the greenhouse effect for a given stellar flux
(Jones et al., 2006). Thus, the distances at both the inner and the
outer HZ boundaries are closer to host star, with higher $T_{\rm
eff}$, than they would have been if the $T_{\rm eff}$ effect is not
taken into consideration.

However, others pointed out that life existence not only needs
clement temperature, but also appropriate ultraviolet radiation
(eg., Setlow \& Doyle, 1954; Lindberg \& Horneck, 1991; Cockell,
1998; Hoyle \& Wickrasinghe, 2003; Sequra et al., 2003). UV
radiation can induce DNA destruction and make life inactivate
(Buccino et al., 2006; Tepfer \& Leach, 2006). And UV radiation is
also one of the most important energy source for the synthesis of
many biochemical compounds on the primitive Earth (Buccino et al.,
2006).

The ``Principle of Mediocrity" is in the ``hard core" of all the
research programs that search for life in the universe (Lakatos,
1974). In the points of this hypothesis, life and intelligence will
develop with the same rules of natural selection wherever the proper
conditions and the needed time are given (von Hoerner, 1961, 1973).
In other words, the conditions that give place to the origin and
evolution of life on Earth are average, in comparison to other
worlds in the universe (Buccino et al., 2006). Using the ``Principle
of Mediocrity", Buccino et al. (2006) gave the boundary UV radiation
of ultraviolet habitable zone (UV-HZ), but not the continuous
UV-ZHs.

Previously, it was pay attention to the HZs of host stars at main
sequence (MS) phase, as the evolution from biochemical compounds to
primary life needs very long time. However, life-seeds may migrate
from one planet to another (Buccino et al., 2007). One hundred Myr
may be too short for the pre-biological evolution, but the features
of biology can change greatly in the same period, based on the
``Principle of Mediocrity". Hence, it is meaningful to study the HZs
and the UV-HZs of host stars at post-MS phase (eg., Franck et al.,
2000; Noble et al., 2002).

Hot subdwarfs are known as Extreme Horizontal Branch stars and
believed to be core He-burning objects with extremely thin hydrogen
envelopes (less than 0.02 \mo). They are an important source of
far-UV light in the galaxy and successfully used to explain the
UV-upturn in elliptical galaxies (Kilkenny et al., 1997; Han et al.
2007). The typical core mass of a hot subdwarf is 0.475 \mo, which
can stably burn more than 160 Myr.

Using the boundary UV radiation of UV-HZ (Buccino et al., 2006), we
achieve the the distances at both the inner and the outer UV-HZ
boundaries, as a function of stellar radius and $T_{\rm eff}$. Using
the data of stellar radius and $T_{\rm eff}$ (Guo et al., 2009,
hereafter Paper I), we calculate the UV-HZs around host stars with
masses less than 4.00 \mo, at zero age main sequence (ZAMS) and at
the terminal of main sequence (TMS). Comparing the UV-HZs with the
HZs (calculated in Paper I) of the same host stars, we find that the
UV-HZs are near to the HZs for the host stars with $T_{\rm eff}$
from 4,600 to 7,137 K. For the host stars with $T_{\rm eff}$ lower
than 4,600 K, the UV-HZs are closer than the HZs, which means that
there is inadequate UV radiation in the HZs. For the host stars with
$T_{\rm eff}$ higher than 7,137 K, the UV-HZs are farther than the
HZs, which means that there is too strong UV radiation and DNA may
be destructed in the HZs.

Using the evolutionary data of a hot subdwarf calculated by Zhang et
al. (2009), we obtain both the HZ and the UV-HZ of the hot subdwarf
with core mass 0.475 \mob and envelope mass 0.001 \mo. It is found
that the UV-HZ is about ten times farther than the HZ for the hot
subdwarf. This means that the UV radiation in the HZ is about one
hundred times stronger than the UV radiation suited to life
existence. Therefore, there is no chance that life survive in the
HZs of hot subdwarfs, for the damaging UV radiation.

The outline of the paper is as follows: we describe our methods in
Section 2, show our results in Section 3, present some discussions
in Section 4, and then finally in Section 5 we give our conclusions.
\section{Methods}
\subsection{Input physics about stellar evolution}
We use the stellar evolution code of Eggleton (1971, 1972, 1973),
which has been updated with the latest input physics over the last
three decades (Han et al., 1994; Pols et al., 1995, 1998). We set
the convective overshooting parameter, $\delta_{\rm OV}=0.12$ (Pols
et al., 1997; Schr\"{o}der et al., 1997). In our calculation, the
value of metallicity is 0.02 and stellar mass is from 0.10 to 4.00
\mob (Paper I).

We adopt the metal mixture by Grevesse \& Sauval (1998). We use OPAL
high temperatures opacity tables (Iglesias \& Rogers, 1996; Eldridge
\& Tout, 2004) in the range of
$4.00<\mathrm{log(}\mathit{T}\mathrm{/K)}\leq 8.70$, and the new
Wichita state low temperature molecular opacity tables (Ferguson et
al., 2005) in the range of
$3.00\leq\mathrm{log(}\mathit{T}\mathrm{/K)}\leq 4.00$. And we have
made the opacity tables match well with Eggleton's code (Chen \&
Tout, 2007; Guo et al., 2008).
\subsection{Stellar luminosity, Radius and $T_{\rm eff}$}
In the subsection 3.1 of Paper I, we gave the fitting formulae of
luminosities and radiuses of stars with masses from 0.08 to 4.00
\mo, at ZAMS and at TMS. And stellar $T_{\rm eff}$ can be obtained
from $L=4{\pi}R^{2}{\sigma}T_{\rm eff}^{4}$. Thus, we can obtain
luminosities, radiuses and $T_{\rm eff}$ of host stars, which can be
used to calculate the HZs and the UV-HZs of the host stars.
\subsection{Boundary flux of HZ}
The inner HZ boundary is determined by the loss of water via
photolysis and hydrogen escape. And the outer HZ boundary is
determined by the condensation of $\rm CO_2$ crystals out of the
atmosphere (von Bloh et al., 2007). Jones et al. (2006) gave the
flux at both the inner and the outer HZ boundaries, as a function of
$T_{\rm eff}$.
\begin{equation}
\frac{S_{\rm in}}{S_{\odot}}=4.190\times10^{-8}T_{\rm
eff}^{2}-2.139\times10^{-4}T_{\rm eff}+1.296, \label{mdis}
\end{equation}
\vspace{-5.0mm}
\begin{equation}
\frac{S_{\rm out}}{S_{\odot}}=6.190\times10^{-9}T_{\rm
eff}^{2}-1.319\times10^{-5}T_{\rm eff}+0.2341, \label{mdis}
\end{equation}
where $S_{\odot}$ is solar constant and $T_{\rm eff}$ is in Kelvin.
\subsection{Boundary UV radiation of UV-HZ}
UV radiation is a double-edged sword to life. If it is too strong,
the terrestrial biological systems will be damaged. And if it is too
weak, the synthesis of many biochemical compounds can not go along.
Therefore, UV radiation should fitly not damage DNA at the inner
UV-HZ boundary, and nicely supply enough energy for the synthesis of
biochemical compounds at the outer UV-HZ boundary. Buccino et al.
(2006) gave the expression of UV photons at both the inner and the
outer UV-HZ boundaries:
\begin{equation}
N_{\mathrm{in}}=\int\limits_{200nm}^{315nm}
B(\lambda)\frac{\lambda}{hc}\frac{L_{\lambda}}{4\pi d^{2}}
\mathrm{d}\lambda, \label{mdis}
\end{equation}
\begin{equation}
N_{\mathrm{out}}=\int\limits_{200nm}^{315nm}
\frac{\lambda}{hc}\frac{L_{\lambda}}{4\pi d^{2}} \mathrm{d}\lambda.
\label{mdis}
\end{equation}
Where $L_{\lambda}$ is stellar luminosity at wave length $\lambda$,
and $B(\lambda)$ is the probability of a UV photon of energy
$(hc)/\lambda$ to dissociate free DNA, whose expression is
\begin{equation}
\mathrm{log}B(\lambda)\sim\frac{6.113}{1+\mathrm{exp}((\lambda[\mathrm{nm}]-310.9)/8.8)}-4.6.
\label{mdis}
\end{equation}

Primary life had come forth on the Archean Earth about 3.8 Gyr ago
(Rosing, 1999). Based on the ``Principle of Mediocrity", a
terrestrial planet needs to receive form half to two times of the UV
radiation received by the Archean Earth, to be suited to biological
evolution (Buccino et al., 2006). The expressions are just as
\begin{equation}
N_{\mathrm{in}}=2N_{\mathrm{in}}(\mathrm{Arc}), \label{mdis}
\end{equation}
\begin{equation}
N_{\mathrm{out}}=0.5N_{\mathrm{out}}(\mathrm{Arc}). \label{mdis}
\end{equation}
\section{Results}
\subsection{Boundary distances of UV-HZs}
We simply take that star is a black body, the expression of
$L_{\lambda}$ is
\begin{equation}
L_{\lambda}=4\pi R^{2}\pi
\frac{2hc}{\lambda^{3}}\frac{1}{e^{hc/kT_{\mathrm{eff}}\lambda}-1}.
\label{mdis}
\end{equation}
Combining Eqs. (3), (6) and (8), it obtains the expression of the
distance at the inner UV-HZ boundary:
\begin{equation}
d_{\mathrm{in}}=\frac{\sqrt{2}}{2}\frac{R}{R_{\mathrm{Arc}}}\sqrt{\frac{F_{1}(T_{\mathrm{eff}})}{F_{1}(T_{\mathrm{Arc}})}}.
\label{mdis}
\end{equation}
Where $R$ is in solar units, $T_{\rm eff}$ is in Kelvin and
$d_{\mathrm{in}}$ is in units of AU. The values of
$R_{\mathrm{Arc}}$ and $T_{\mathrm{Arc}}$ are 0.9113 $R_{\odot}$ and
5,603 K, just as the radius and the $T_{\rm eff}$ of Solar about 3.8
Gyr ago, respectively. The expression of $F_{1}$ is
\begin{equation}
F_{1}(T)=\int\limits_{200nm}^{315nm}
\frac{B(\lambda)}{\lambda^2}\frac{\mathrm{d}\lambda}{e^{hc/kT
\lambda}-1}. \label{mdis}
\end{equation}

Combining Eqs. (4), (7) and (8), it achieves the expression of the
distance at the outer UV-HZ boundary:
\begin{equation}
d_{\mathrm{out}}=\sqrt{2}\frac{R}{R_{\mathrm{Arc}}}\sqrt{\frac{F_{2}(T_{\mathrm{eff}})}{F_{2}(T_{\mathrm{Arc}})}}.
\label{mdis}
\end{equation}
And the expression of $F_{2}$ is
\begin{equation}
F_{2}(T)=\int\limits_{200nm}^{315nm}
\frac{1}{\lambda^2}\frac{\mathrm{d}\lambda}{e^{hc/kT \lambda}-1}.
\label{mdis}
\end{equation}
\subsection{HZs and UV-HZs around host stars}
According to Eqs (9)-(12) and the correlative fitting formulae in
the subsection 3.1 of Paper I, we calculate the UV-HZs around host
stars with masses from 0.08 to 4.00 \mob at ZAMS. As the MS
lifetimes of M type stars are from 131 Gyr to several trillion
years, which are many times longer than the universe age. Hence, M
type stars with masses from 0.08 to 0.50 \mob almost stay at ZAMS,
within the universe age. Therefore, we only calculate the UV-HZs
around host stars with masses from 0.50 to 4.00 \mob at TMS.

As we have given the HZs around host stars with masses from 0.08 to
4.00 \mob (Paper I). In order to comparing the UV-HZs with the HZs
of host stars more intuitively, we put them on the same graphs, seen
in Figs. 1 and 2. It is seen that the UV-HZs are near to the HZs for
solar-like stars, the UV-HZs are closer than the HZs for M and K
type stars, and the UV-HZs are farther than the HZs for upper MS
stars.
\begin{figure}
\psfig{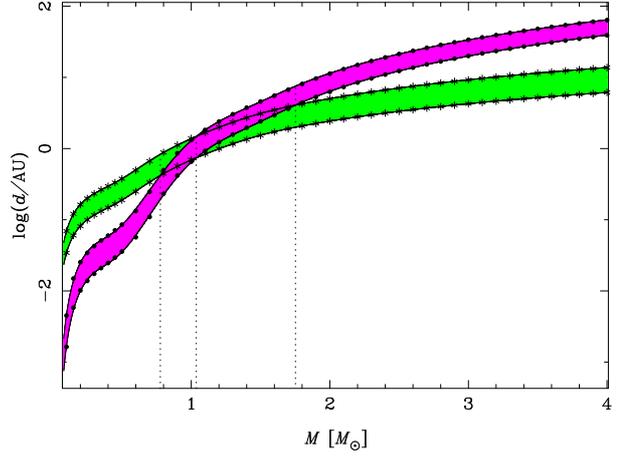}
\caption{HZs and UV-HZs of host stars at ZAMS. The lines with
asterisks denote the boundary distances of HZs, and the lines with
points denote the boundary distances of UV-HZs. Points and asterisks
are stellar data taken from Eggleton's code, and the lines are taken
from fitting formulae, the same as Fig. 2.} \label{ised}
\end{figure}
\begin{figure}
\psfig{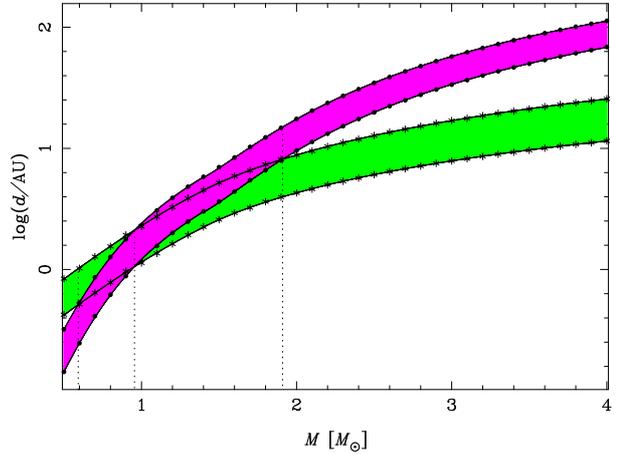}
\caption{HZs and UV-HZs of host stars at TMS.} \label{ised}
\end{figure}

This is because both the boundary flux of HZ and the boundary UV
radiation of UV-HZ depend on $T_{\rm eff}$ for a given stellar flux,
but with the contrary effects. For HZ, the higher $T_{\rm eff}$, the
less the infrared fraction in luminosity, and the less this
fraction, the less the greenhouse effect for a given stellar flux.
Therefore, a terrestrial planet around a host star with higher
$T_{\rm eff}$ needs greater flux to remain the water in liquid
phase. Thus, the distances at both the inner and the outer HZ
boundaries are closer to the host stars with higher $T_{\rm eff}$,
and farther to the host stars with lower $T_{\rm eff}$, for a given
stellar flux.

On the contrary, the higher $T_{\rm eff}$, the higher the UV
fraction in luminosity. Therefore, a terrestrial planet around a
host star with higher $T_{\rm eff}$ need lower flux to be suitable
for life existence. Thus, the distances at both the inner and the
outer UV-HZ boundaries are farther to the host stars with higher
$T_{\rm eff}$, and closer to the host stars with lower $T_{\rm
eff}$, for a given stellar flux. Hence, the UV-HZs are farther than
the HZs for the host stars with higher $T_{\rm eff}$, and closer
than the HZs for the host stars with lower $T_{\rm eff}$.

For the host stars with masses less than 0.777 \mob at ZAMS and the
host stars with masses less than 0.590 \mob at TMS, the UV-HZs at
the outer boundary are closer than the HZs at the inner boundary.
And the host star with mass 0.777 \mob at ZAMS (see the left dotted
line of Fig. 1) and the host star with mass 0.590 \mob at TMS (see
the left dotted line of Fig. 2) have the same $T_{\rm eff}$, just as
4,600 K. Hence, the UV-HZs are closer than the HZs and there is no
intersection between them, for the host stars with $T_{\rm eff}$
lower than 4,600 K. And the lower are the $T_{\rm eff}$ of host
stars, the more are the differences between the UV-HZs and the HZs
of the host stars.

For the host stars with masses more than 1.752 \mob at ZAMS and the
host stars with masses more than 1.910 \mob at TMS, the UV-HZs at
the inner boundary are farther than the HZs at the outer boundary.
And the host star with mass 1.752 \mob at ZAMS (see the right dotted
line of Fig. 1) and the host star with mass 1.910 \mob at TMS (see
the right dotted line of Fig. 2) have the same $T_{\rm eff}$, just
as 7,137 K. Hence, the UV-HZs are farther than the HZs and there is
no intersection between them, for the host stars with $T_{\rm eff}$
higher than 7,137 K. And the higher are the $T_{\rm eff}$ of host
stars, the more are the differences between the UV-HZs and the HZs
of the host stars.

Therefore, the UV-HZs and the HZs are near to each other and there
are intersections between them, for the host stars with $T_{\rm
eff}$ from 4,600 to 7,137 K. For the host star with mass 1.037 \mob
at ZAMS (see the middle dotted line of Fig. 1) and the host star
with mass 0.955 \mob at TMS (see the middle dotted line of Fig. 2),
the UV-HZs and the HZs entirely coincide with each other, at both
the inner and the outer boundaries. And the host star with mass
1.037 \mob at ZAMS and the host star with mass 0.955 \mob at TMS
have the same $T_{\rm eff}$, just as 5,636 K. Hence, the UV-HZs and
the HZs are in the same regions, for the host stars with $T_{\rm
eff}$ 5,636 K.
\subsection{Impacts of the differences between HZs and UV-HZs on biological evolution}
For the host stars with $T_{\rm eff}$ from 4,600 to 7,137 K, most of
which are solar-like stars, the UV-HZs are near to the HZs. This
means that there is appropriate UV radiation in the HZs for
solar-like stars, which is suited to biological evolution.

However, the UV-HZs are closer than the HZs for the host stars with
$T_{\rm eff}$ lower than 4,600 K, most of which are M and K type
stars. This means that there are inadequate UV radiation in the HZs
for M and K type stars, especially for M type stars. And UV
radiation is used to drive the synthesis of many essential
biomolecules. Fortunately, stellar flares of M type stars can
generate adequate UV radiation (Heath et al., 1999), which supply
the energy source for the synthesis of many biochemical compound.
But the stellar flares had better not be too strong, which may
damage biological systems. Hence, M type stars with moderate flares
are the best candidates to host habitable planets (Buccino et al.,
2007).

On the contrary, the UV-HZs are farther than the HZs for the host
stars with $T_{\rm eff}$ higher than 7,137 K, most of which are
upper MS stars. This means that there are too strong UV radiation in
the HZs for upper MS stars, which can induce DNA destruction and
cause damage to a wide variety of proteins and lipids. Therefore,
the probability of life existence around these stars will decrease
dramatically.

For example, the $T_{\rm eff}$ is 10,473 K at ZAMS, for the host
star with mass 3.00 \mo. The distance at the inner UV-HZ boundary is
20.008 AU, where the UV radiation nicely does not destory DNA. And
the distances at both the inner and the outer HZ boundaries are
4.254 and 9.236 AU, respectively. Therefore, the UV radiation in the
HZ is from 4.693 to 22.117 times of the UV radiation damaging DNA,
which is not suited to life existence.
\subsection{HZ and UV-HZ of hot subdwarf}
Hot subdwarfs are known as Extreme Horizontal Branch stars and
believed to be core He-burning objects with extremely thin hydrogen
envelopes. In this paper, the core mass of hot subdwarf is a typical
value 0.475 \mob and envelope mass is 0.001 \mo, which is calculated
by Zhang et al. (2009). And we obtain both the HZ and the UV-HZ of
the hot subdwarf at the whole evolutionary phase, seen in Fig. 3.
\begin{figure}
\psfig{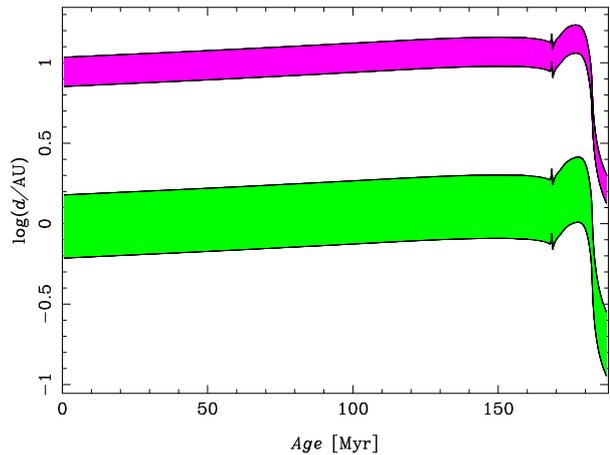}
\caption{HZ (lower zone) and UV-HZ (upper zone) of hot subdwarf.}
\label{ised}
\end{figure}

For example, the $T_{\rm eff}$ of the hot subdwarf is 33,029 K, with
age 50 Myr. The distance at the inner UV-HZ boundary is 7.838 AU.
And the distances at both the inner and the outer HZ boundaries are
0.673 and 1.661 AU, respectively. Therefore, the UV radiation in the
HZ is from 22.257 to 135.696 times of the UV radiation destructing
DNA, and this strong UV radiation can easily kill the lives living
in the HZ.
\section{Discussion}
In this paper, we use the boundary UV radiation of Buccino et al.
(2007). In the UV-HZ of a host star, the number of UV photons $N$
meets $0.5N_{\rm Arc}\le N\le 2N_{\rm Arc}$. This confine may be too
strict, how about $0.25N_{\rm Arc}\le N\le 4N_{\rm Arc}$? That may
induce broader UV-HZ, and the intersection between UV-HZ and HZ will
enlarge. However, the UV-HZs are also closer than the HZs for the
host stars with lower $T_{\rm eff}$, and farther than the HZs for
the host stars with higher $T_{\rm eff}$.

X-ray and extreme ultraviolet (EUV) could also play a key role in
the origin and development of life on Earth and possibly on Mars
(Luhmann \& Bauer, 1992). Using the ``Principle of Mediocrity", we
can also calculate X-ray habitable zones and EUV habitable zones of
host stars. It is speculated about that these two habitable zones
are also closer than the HZs for the host stars with lower $T_{\rm
eff}$ and farther than the HZs for the host stars with higher
$T_{\rm eff}$. And the intersection between these two habitable
zones and HZ for the same host star will decrease.
\section{Conclusion}
Firstly, we give the boundary distances of UV-HZ, as a function of
stellar radius and $T_{\rm eff}$. we obtain the UV-HZs around host
stars with masses from 0.08 to 4.00 \mo, and compare the UV-HZs with
the HZs. The UV-HZs are closer than the HZs for the host stars with
$T_{\rm eff}$ lower than 4,600 K, and farther than the HZs for the
host stars with $T_{\rm eff}$ higher than 7,137 K. Secondly, we make
out the impacts of the differences between HZs and UV-HZs on
biological evolution. The UV radiation is very strong in the HZs for
upper MS stars, which is negative to live existence. Thirdly, we
give the HZ and the UV-HZ of a hot subdwarf, with core mass 0.475
\mob and envelope mass 0.001 \mo. The UV radiation in the HZ is from
22.257 to 135.696 times of the UV radiation dissociating DNA, which
can easily kill lives in the HZ. Finally, we present discussions
about the boundary UV radiation and X-ray and EUV radiation. One may
also send any special request to \it{guojianpo1982@hotmail.com}
\normalfont{or} \it{guojianpo16@163.com}\normalfont{.}
\begin{acknowledgements}
This work is supported by the National Natural Science Foundation of
China (Grant Nos. 10773026, 10821061 and 2007CB815406), the Chinese
Academy of Sciences (Grant No. KJCX2-YW-T24) and Yunnan Natural
Science Foundation (Grant No. 06GJ061001).
\end{acknowledgements}

{}
\end{document}